\documentclass[twocolumn,english,showpacs,nofootinbib]{revtex4-1}
\usepackage[T1]{fontenc}
\usepackage[latin9]{inputenc}
\usepackage{geometry}
\usepackage{xcolor}
\usepackage{babel}
\usepackage{units}
\usepackage{amsmath}
\usepackage{amssymb}
\usepackage{stmaryrd}
\usepackage{graphicx}
\usepackage{wasysym}
\usepackage[bookmarks=false]{hyperref}

\makeatletter

\IfFileExists{lmodern.sty}{\usepackage{lmodern}}{}
\usepackage{babel}

\usepackage{hyperref}

\makeatother

\begin{document}
\title{Quantum-Logic Detection of Chemical Reactions}

\author{Or Katz}
\thanks{Present address: Department of Electrical and Computer Engineering, Duke University, Durham, North Carolina 27708, USA}
\email{Corresponding author: or.katz@duke.edu}
\affiliation{Department of Physics of Complex Systems, Weizmann Institute of Science, Rehovot 7610001, Israel}

\author{Meirav Pinkas}
\affiliation{Department of Physics of Complex Systems, Weizmann Institute of Science,
Rehovot 7610001, Israel}

\author{Nitzan Akerman}
\affiliation{Department of Physics of Complex Systems, Weizmann Institute of Science,
Rehovot 7610001, Israel}

\author{Roee Ozeri}

\affiliation{Department of Physics of Complex Systems, Weizmann Institute of Science,
Rehovot 7610001, Israel}

\begin{abstract}
Studies of chemical reactions by a single pair of atoms in a well defined quantum state constitute a corner stone in quantum chemistry. Yet, the number of demonstrated techniques which enable observation and control of a single chemical reaction is handful. Here we propose and demonstrate a new technique to study chemical reactions between an ultracold neutral atom and a cold ion using quantum logic. We experimentally study the release of hyperfine energy in a reaction between an ultracold rubidium atom and isotopes of singly ionized strontium for which we do not have experimental control. We detect the reaction outcome and measure the reaction rate of the chemistry ion by reading the motional state of a logic ion via quantum logic, in a single shot. Our work opens new avenues and extends the toolbox of studying chemical reactions, with existing experimental tools, for all atomic and molecular ions in which direct laser cooling and state detection are unavailable.
\end{abstract}
\maketitle

\section*{Introduction}

Chemical reactions have long captured great scientific and applicable interest. When two reactants are brought into close proximity, they can change their physical state and convert their internal energy into an increase or decrease of their kinetic energy. Traditionally, chemical reactions are studied by measuring variations in average thermodynamic properties such as temperature, energy and pressure of macroscopic samples comprising many atoms. 

In the microscopic scale, the dynamics and reaction outcome are governed by the rules of quantum mechanics. Short-range exchange forces typically produce many-body correlations in the electronic state of the reactants, which comprise a molecular complex \cite{qchem}. As modest molecular complexes typically consist of tens to hundred electrons, and as the necessary computing power for an exact ab-initio computation scales exponentially with the number of electrons, accurate calculation of a single reaction outcome yet remains a great computational challenge \cite{CompPow,CompPow2}. Consequently, accurate theoretical modeling often requires an experimental calibration of several free parameters.

The advent of ultra-cold atomic gasses enabled the study of chemical reactions in the ultra-cold regime. In this regime collisions and reactions are governed by quantum dynamics. Here, resonant scattering phenomena, such as Feshbach and shape resonances, enable the coherent and efficient reaction between atoms \cite{atoms_Feshbach,atoms_Feshbach2,atoms_shape}. The observation of these resonances has also enabled a precise calibration of molecular potentials.  

Hybrid systems of laser-cooled trapped ions and ultracold neutral atoms offer pristine experimental tools for the study reactions of a single ion-atom pair \cite{hybrid_RMP,Nature_Kohl,hybrid_RMP2,hybrid_RMP3,Ziv_sys}. These systems enable to explore cold reactions as ions can be laser-cooled near their motional ground state and atoms to temperatures below $\mu$K. Various inelastic processes were studied with these systems including spin-exchange \cite{spin_exchange1,spin_exchange2,spin_exchange3,optical_trap}, spin relaxation \cite{spin_relaxation1,spin_relaxation2}, charge-exchange \cite{charge_exchange0,charge_exchange1,charge_exchange2,charge_exchange3,charge_exchange4,charge_exchange5,charge_exchange6}, molecular association and dissociation \cite{Molecular_formation1,Molecular_formation2}, as well as elastic processes \cite{elastic1,elastic2} and nonequilibrium open-system dynamics \cite{nonequilib_1,nonequilib_2,nonequilib_3,nonequilib_4}. While hybrid atom-ion systems provide exquisite control over the physical state of both atom and ion, the necessary degree of control in their preparation and detection often severely limits the range of investigated atomic and molecular species as well as the internal states involved. 

\begin{figure*}[t]
\begin{centering}
\includegraphics[width=17.5cm]{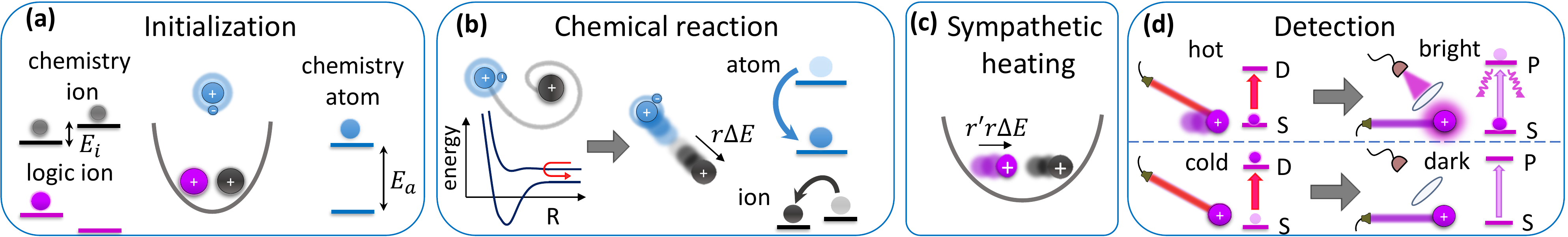}
\par\end{centering}
\centering{}\caption{\textbf{Logic detection of exothermic reactions.} (a) State initialization. The system consists of three bodies: a logic ion co-trapped with a chemistry ion, and a free neutral atom or molecule with which the chemistry ion reacts. The motional and internal states of the logic ion and the atoms can be initialized and coherently controlled (e.g.~via cooling and optical pumping). (b) Chemical reaction. Long range attraction forces lead to spiraling of the atom and chemistry ion, resulting in a cold Langevin collision. The state of the chemistry ion following the collision is determined by the short-range interaction with the atom. In exothermic reactions, scattering between molecular potentials results in a decrease of the total internal energy (e.g.~$\Delta E=E_a+E_i$) and an increase of the kinetic energy of the reactants. Here, the chemistry ion gains a fraction $r=m_a/(m_a+m_i)$ of that energy. The effect of the presence of the logic ion on the atom-chemistry ion scattering is negligible. (c) Sympathetic heating. Strong Coulomb forces between the two ions correlate their motion, as they share the same phononic modes of the trap. The kinetic energy of the chemistry ion is consequently distributed between these modes. For near-equal ion masses the energy is approximately equally distributed on average, such that each mode gains $r'\approx1/6$ of  the energy. (d). Logic detection of motion. The occurrence of a reaction is detected through the motional state of the logic ion. Different state of motion (hot/cold) of the logic ion are mapped onto different internal states (e.g. S or D electronically excited-states) that are detected via state-selective fluorescence (bright or dark).\label{fig:Logic_scheme}}
\end{figure*}

Quantum logic techniques can alleviate these experimental limitations, and enable the preparation and measurement of atomic and molecular species whose access is inefficient or challenging \cite{q_logic1}. These methods are widely applied in spectroscopy of atoms and molecules \cite{q_logic2,q_logic3,q_logic4,q_logic5}, in precision measurements \cite{q_logic_precision,q_logic_precision2} in qunatum information \cite{Leibfried1} and in search of new physics beyond the Standard Model \cite{q_logic_SM}. For trapped ions system, a typical realization consists of a pair of different species, whose motion is coupled by the strong coulomb force. One (logic) ion can be initialized, manipulated and detected by optical means, efficiently measuring the state of the other (spectroscopy) ion by its imprint on the collective motion of the crystal. Several works employed or proposed elements of quantum logic to detect chemical reactions. Charge exchange rates between an ensemble of molecular ions and ensemble of ultra-cold atoms were measured using sympathetic cooling and logic mass spectroscopy via co-trapped ions \cite{willitsch1}. Another work traced molecular association of a single cold molecular ion with room-temperature gas using quantum logic \cite{willitsch1b}. Other recent proposals suggested to employ variation of quantum logic for coherent manipulation of the molecule using phonon resolved transitions \cite{willitsch2}, and to study the Al$^+$ optical clock transition  during its interaction with ultracold atoms \cite{q_logic_chem}. Yet, to our knowledge, quantum logic has never been applied to measure the reaction rate of a single pair of cold reactants whose optical control is inaccessible. 

Here we present and experimentally demonstrate a quantum logic method to study endothermic or exothermic chemical reactions of a single pair of atoms. Our technique enables to measure the reaction of a chemistry ion in a single-shot with high efficiency, via its effect on the motion of another logic ion. We demonstrate the technique via measurement of cold hyperfine-changing exothermic reactions between a $^{87}$Rb atom and a singly-ionized Sr ion. We explore the interaction for all four stable isotopes ($^{84}\text{Sr}^{+}$,$^{86}\text{Sr}^{+}$,$^{87}\text{Sr}^{+}$ and $^{88}\text{Sr}^{+}$) as a chemistry ion using additional $^{88}\text{Sr}^{+}$ as a logic ion. This method opens new avenues for the study of quantum chemistry, with potential applications in measurement of reaction cross-sections, quantum resonances, collisional cooling of atoms and molecules, and reactions and dynamics of optically-inaccessible reactants.

\section*{Results}
\subsection*{Quantum logic technique}
To describe the quantum logic technique we consider a two-ion crystal composed of a logic ion which can be manipulated and detected efficiently, and a chemistry ion whose interaction with background atoms we would like to study. Our method is equally applicable for the study of reactions in which energy is released to, or removed from, the motion of the reactants. Without loss of generality, here we focus on exothermic reactions. 

For an exothermic reaction to occur, we assume that at least one of the two reactants (in our case the atom) is initially prepared in an excited state with a nonzero internal energy $E_a>0$, and that the chemistry ion has an internal energy $E_i$ as presented in  Fig.~\ref{fig:Logic_scheme}(a). Here we assume that the internal states of the logic and neutral atoms can be efficiently initialized, that the atom is laser-cooled, and that the ions are trapped and cooled near the motional ground state. 

When the neutral atom approaches the chemistry ion at a distance $R$, it is attracted by its long-rage universal $-1/R^4$ polarization potential, and spirals inwards towards a short-range (Langevin) collision \cite{cetina_heating} as shown in Fig.~\ref{fig:Logic_scheme}(b). During the collision, short-range exchange forces, which are specific to the interacting atom-ion pair, correlate the quantum state of the complex on different molecular potential energy curves \cite{atom_ion_cote}. For the current discussion, we assume that the molecular complex dissociates following the collision. The final state of the colliding parties depends on their initial states and the scattering amplitudes of the reaction. 
As energy is conserved, the internal energy difference between the initial and final states of the colliding atoms, $\Delta E=E_a\pm E_i$, is converted into kinetic energy. In the center of mass frame an energy of $r\Delta E$ is released into the kinetic motion of the chemistry ion where $r=m_a/(m_a+m_i)$, and $m_a$ and $m_i$ are the masses of the chemistry atom and ion respectively. 

The energy release into the motion of the chemistry ion sympathetically stimulates the motion of the mutually trapped logic ion, as shown in Fig.~\ref{fig:Logic_scheme}(c). The energy is distributed between the six motional modes of the trap. For nearly equal ion masses the energy is, on average, equally divided between the modes. Detection of the motion of the logic ion can be realized via various optical thermometry techniques which rely on state-dependent fluorescence and the Doppler effect \cite{Ozeri_2007,nonequilib_1}. The particular detection technique realized in our experiment and shown in Fig.~\ref{fig:Logic_scheme}(d), enables single-shot detection of the reaction using electron-shelving as outlined below.

\subsection*{Experimetnal implementation}
\begin{figure}[b]
\begin{centering}
\includegraphics[width=8.6cm]{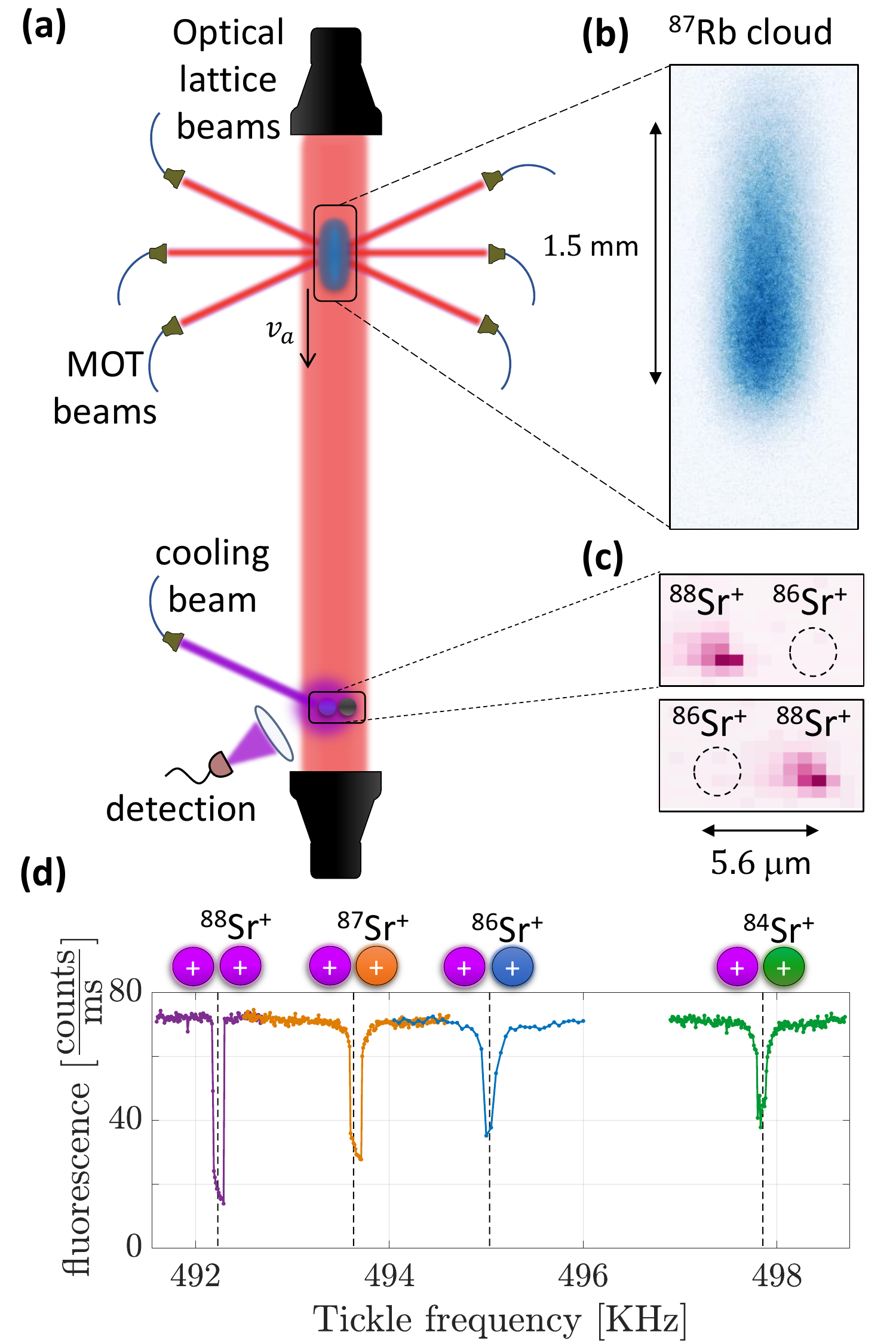}
\par\end{centering}
\centering{}\caption{\textbf{Experimental setup.} (a) Setup schematics. A cloud of neutral atoms is laser cooled and loaded into an off-resonant optical dipole trap. A pair of ions is loaded in a linear Paul and cooled near the ground state. The dilute atomic cloud is shuttled through the ions trap, such that on average up to a single spiraling collision with one of the ions occurs. (b) Absorption imaging of the atomic cloud in the top chamber. (c) Fluorescence imaging of a two ion crystal composed of a logic ion ($^{88}$Sr$^{+}$) and a chemistry ion ($^{86}$Sr$^{+}$) which is always dark. The two orderings of the ions crystal are shown to exemplify the presence of the dark ion. (d) Verification of the chemistry ion isotope. We selectively load a particular isotope, and verify its mass via measurement of the resonance frequency of the center of mass mode. The latter is inferred via decrease of the logic ion's fluorescence to resonant time-dependent external electric field ("tickle"). \label{fig:setup}}
\end{figure}

The experimental setup consists of a mixture of laser-cooled atoms and electrically trapped ions \cite{Ziv_sys}, as shown schematically in Fig.~\ref{fig:setup}(a). We collect and laser cool a Magneto-Optical Trap (MOT) of $10^6$ $^{87}$Rb atoms that are efficiently loaded into an off-resonant optical lattice. An absorption image of the trapped atoms is shown in Fig.~\ref{fig:setup}(b). We control the internal spin state of the atoms in the ground state, $|F,m\rangle$, using a sequence of microwave and optical-pumping pulses. We then shuttle the cloud to the lower vacuum chamber by varying the relative optical frequencies of two counter-propagating optical lattice beams.  The quantum numbers of the total atomic spin  of the rubidium atoms, and their projection along the magnetic field axis are $F$ and $m$ respectively.

In addition, we trap a two-ion crystal in a linear Paul trap located in a lower vacuum chamber. The crystal is composed of $^{88}\text{Sr}^{+}$ as the logic ion (purple) and one of the stable isotopes $^{84}\text{Sr}^{+}$,$^{86}\text{Sr}^{+}$,$^{87}\text{Sr}^{+}$ or $^{88}\text{Sr}^{+}$ as a chemistry ion (black). We realize isotope-selective loading via frequency tuning of the photo-ionizing laser. In Fig.~\ref{fig:setup}(c) we present fluorescence imaging of the two ions, where the bright logic ion efficiently scatters photons of the detection laser, whereas the chemistry ion ($^{86}\text{Sr}^{+}$), is considerably off-resonant and therefore does not appear. We identify the isotope of the chemistry ion via mass spectrometery \cite{Drewsen1} as shown in Fig.~\ref{fig:setup}(d). Here, we apply resonant electric fields ("tickle") that heat the mass-dependent axial-motional mode of the crystal, and monitor the resulting decrease in the florescence of the logic ion due to heating. Further details on the experimental sequences and apparatus are given in the Supplementary Material.

During the shuttling of atoms towards the lower chamber, the logic ion is laser cooled near its ground motional state via side-band cooling, and its spin is optically-pumped. The chemistry ion is sympathetically cooled near its motional ground state by its electrostatic coupling to the laser-cooled logic ion, and its spin state is driven into a mixed-state by sporadic absorption of off resonant photons. As the sparse atomic cloud traverses through the ion-trap, an atom can be captured by the attractive long-range polarization potential of one of the two ions into a collision. To study the effect of single collision events, we set the probability of Langevin-type collision of each ion per atomic cloud passage to be relatively small (about $25\%$), by controlling the cloud velocity and the MOT loading time. 

\begin{figure*}[t]
\begin{centering}
\includegraphics[width=17.7cm]{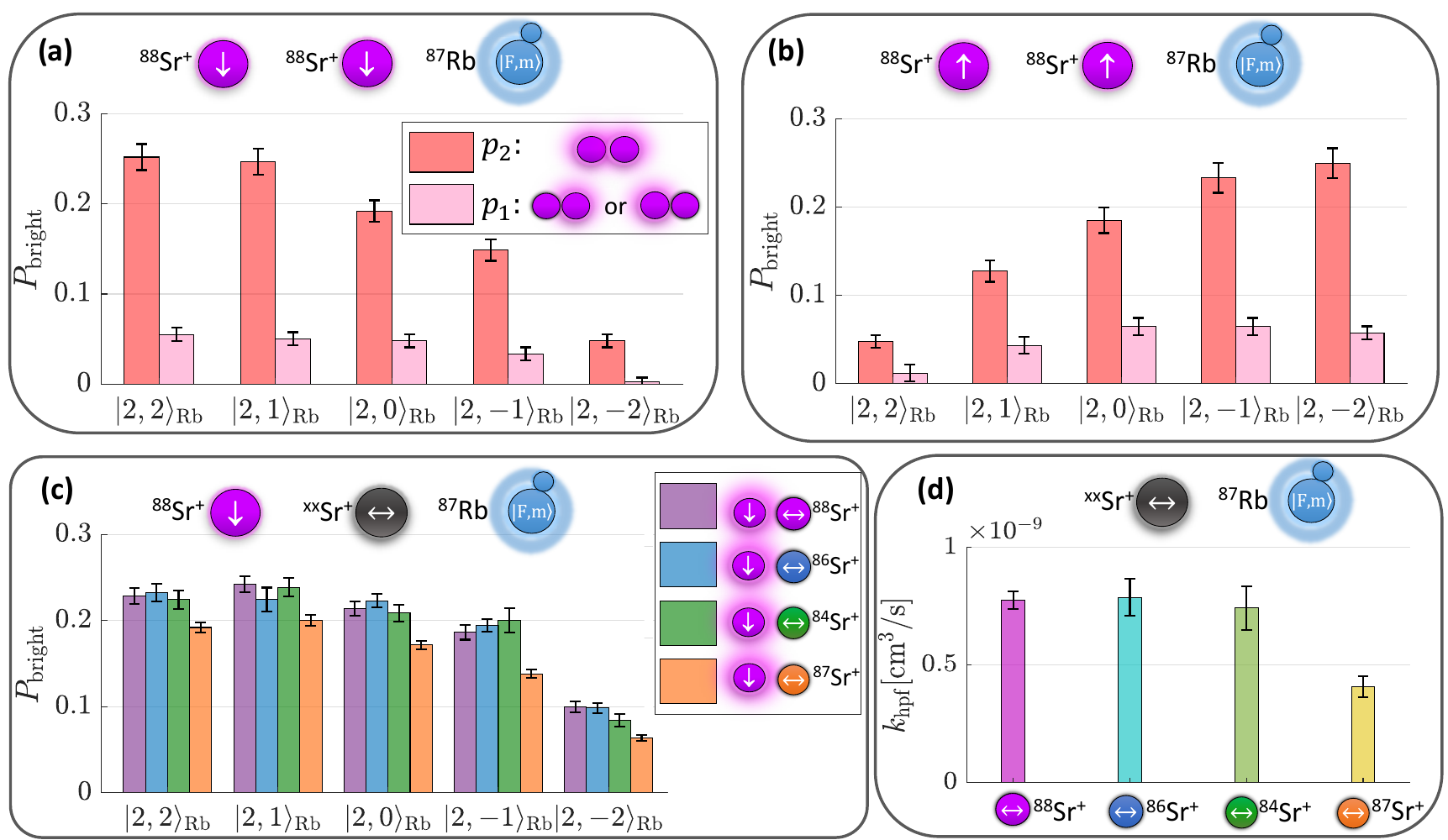}
\par\end{centering}
\centering{}\caption{\textbf{Logic detection of hyperfine-changing reactions.} (a+b) Probability to measure a bright (hot) ion of a two $^{88}\text{Sr}^{+}$ ion crystal for different internal states of the $^{87}$Rb atoms. The probability of correlated events ($p_2$) is significantly higher than the probability of uncorrelated events ($p_1$), indicating high detection efficiency (about $80\%$) of the shelving-detection. Ions are prepared in $|\downarrow\rangle$ (a) or $|\uparrow\rangle$ (b). (c). Probability to measure a single spin-down logic ion $^{88}\text{Sr}^{+}$ bright, and any of the isotopes of Sr$^{+}$ as chemistry ion in an unpolarized state. The statistics include both collision of an atom with the chemistry ion or with the logic ion. (d). The hyperfine-changing reaction rate of the different isotopes. We quantify the reaction rate of the atom and the chemistry ion by subtracting from (c) the collision probability of the logic ion from (a), normalizing by the detection efficiency, and averaging over the Rb internal states. Data in (a-c) excludes SPAM errors and additional non hyperfine-changing stray heating. The exclusion is realized by subtracting from the measured data the small probability of hot events in an independent experiment in which the atoms are initialized in the lower hyperfine manifold ($F=1$). Bars represent $1\sigma$ binomial uncertainties\label{fig:Fig4}}
\end{figure*}

\subsection*{Hyperfine-changing collisions}

At the onset of a collision, the spin-dependent scattering channels of the atom-ion complex are dominated by the bare atomic states, determined by the atomic hyperfine coupling to nuclear spins and Zeeman coupling to the magnetic field. For $^{87}$Rb, the coupling of the electron to the $3/2$ spin in its nucleus sets a large frequency gap between the lower ($F=1$) and upper ($F=2$) hyperfine manifolds of about $6.8$ GHz (i.e.~ an energy gap of $E_a= k_B\times (328$ mK) where $k_B$ is the Boltzmann constant), whereas even isotopes of Sr$^{+}$ have zero nuclear spin and their Zeeman magnetic splitting in the experiment is small ($E_i<k_B\times 1$ 
mK). As the atoms spiral inward to short distances, the two valence electrons of the atom-ion complex experience spin-dependent molecular interaction. The dominant short-range interactions in our experiment are spin-exchange, corresponding to the singlet and triplet molecular curves, and weaker spin-relaxation which couples the electronic spins to the orbital angular momentum of the complex. Both interactions can mix the spin-channels of the isolated atoms, and alter their initial spin state after scattering. 

In this work we focus on hyperfine-changing collisions, in which the rubidium, initially in one of the $|F=2,m\rangle$ states, is scattered to any of the states in the $F=1$ manifold. As the mass of rubidium and strontium is nearly equal ($r\approx1/2$), the chemistry ion typically gains half of the internal energy ($164$ mK) in motion. This energy is distributed between the phononic modes of the two-ion crystal, heating each mode by $27$ mK on average, and consequently setting the logic ion in motion.

\subsection*{Logic detection}

To characterize the reaction rate, we measure the occurrence probability of a single hyperfine-changing reaction by detecting the change in motion of the logic ion. 

Different ion thermometery techniques were previously developed to characterize and detect the motion of trapped-ions. Sideband and carrier Rabi thermometery is mostly sensitive at low ion temperatures below a few milliKelvin per mode, whereas Doppler cooling thermometery is mostly sensitive at temperatures above hundred milliKelvin per mode \cite{Ziv_sys}. However, these techniques are relatively inefficient in single-shot detection of motion of tens milliKelvin per mode as in heating by a single hyperfine-changing collision. Here we realize and numerically characterize a different coherent electron-shelving technique, which maps a hot ion to the S ground state but shelves a cold ion to the D electronically-excited state. 

Our method uses the following operations after the interaction of the ions with the atomic cloud. First, we apply optical pumping pulses to negate the affect of spin-exchange or relaxation of the logic ion spin on the detection. Then, we apply two $\pi$-pulses of a narrowband $674$ nm light, tuned on resonance with two different optical transitions between the $\textrm{S}_{1/2}$ and the $\textrm{D}_{5/2}$ manifolds, as shown in Fig.~\ref{fig:Logic_scheme}(d) for a single transition. When the ion is cold (absent a reaction in which the hyperfine energy was released), we succeed in transferring it to the D state with above $99\%$ fidelity. In contrast, when the logic ion is heated, it oscillates in the trap with large amplitude and modulates its position with respect to the optical wavefront of the shelving beam. The periodic motion compromises the coherent shelving operation, thus lowering the shelving probability and leaving a hot ion in the S state. We then discriminate between the hot (S) and cold (D) ion cases by state selective-fluorescence on the $\textrm{S}_{1/2} \rightarrow \textrm{P}_{1/2}$ strong dipole transition. Numerical characterization of this shelving technique is presented in Methods.

\subsection*{Reaction-rate measurements}
In Fig.~\ref{fig:Fig4} we present the probability that the logic ion appeared bright ($P_{\text{bright}}$) after the atomic cloud passage for different crystal spin configurations. To characterize and account for direct heating of the logic ion by collisions, we first characterize the reaction of $^{87}\text{Rb}$ in any of the magnetic levels within the $F=2$ manifold with a crystal of two $^{88}\text{Sr}^+$ whose spins were initialized either in $|\downarrow\rangle$ (Fig.~\ref{fig:Fig4}a) or $|\uparrow\rangle$ (Fig.~\ref{fig:Fig4}b). As the state of both ions can be efficiently detected, we characterize the probability in which the two ions appeared bright as $p_2$ and the probability that only a single ion appeared bright as $p_1$. The results presented exclude the effect of SPAM errors and stray heating (e.g.~ via RF heating \cite{cetina_heating,Pinkas2020}), by independently measuring the small probability the ion appeared bright when the Rb cloud is initialized in $F=1$ and subtracting this probability from the results we present. Therefore, all bright events in Fig.~\ref{fig:Fig4} are associated with hyperfine-changing reactions with the Rb atom.

It is apparent that reactions in which the atomic and ion spin are anti-parallel are more probable, and that the brightness of the two $^{88}\text{Sr}^+$ ions is correlated ($p_2>p_1$). The former observation agrees with the dominance of the spin-exchange interaction between Sr$^{+}$-Rb, whereas the latter is associated with the strong sympathetic heating of the ions and the relatively high detection efficiency of our scheme, being $\eta=p_2/(p_1+p_2)\approx0.8$. 

In Fig.~\ref{fig:Fig4}c we present the brightness of the logic ion for each of the three isotopes $^{84}$Sr$^{+}$, $^{86}$Sr$^{+}$, and $^{87}$Sr$^{+}$ when the logic ion was prepared in $|\downarrow\rangle$, but the chemistry ion is spin unpolarized (marked by $\leftrightarrow$ in Fig.~\ref{fig:Fig4}c), as dictated by the linear polarization and unresolved spectrum of the cooling light. We additionally present the case of one of the $^{88}$Sr$^{+}$ ions being in a completely-mixed spin-state, constructed by the data in Figs.~\ref{fig:Fig4}(a-b) and also marked by $\leftrightarrow$. Notably, the decrease in the reaction probability for decreasing values of $m$ is a result of the spin-downward orientation of the logic ion. While $^{87}$Sr$^{+}$ has a nonzero nuclear spin ($I=9/2$) and a hyperfine splitting of about $5$ GHz, no excess heating was observed for collisions with Rb atoms prepared in the $F=1$ manifold, with respect to even isotopes of strontium. As a hyperfine change of $^{87}$Sr$^{+}$ could have been observed, this indicates that this ion was optically pumped to the lower $F=5$ hyperfine manifold due to residual scattering of the logic ion's cooling light. 

To estimate the reaction rate of the chemistry ion, we subtract the direct effect of a single logic ion, $(p_1+p_2)/2$, from the data in Fig.~\ref{fig:Fig4}c and normalize by the detection efficiency $\eta$. The hyperfine changing collisions rate of the different chemistry ions are presented in Fig.~\ref{fig:Fig4}d, averaged over the five magnetic state of the Rb atoms in $F=2$. The absolute rate coefficient was determined by normalizing the results with the measured rate coefficient with the maximally polarized channel measured in Ref.~\cite{spin_exchange2} for $^{88}\text{Sr}^{+}$. Evidently, the even isotopes have similar hyperfine-changing rate coefficient but the odd isotope, $^{87}\text{Sr}^+$, has a rate which is almost twice smaller.

\section*{Discussion}
The measured reaction rates can be used to calibrate ab-initio calculations and determine the molecular potentials of $^{87}$Rb with the different isotopes of Sr$^{+}$, independently extracting the spin-exchange and spin-relaxation interaction strengths. Furthermore, previous works suggested that quantum signatures of the ultra-cold s-wave scattering (about a hundred nK for Rb-$^{+}$Sr), would have imprint on the measured reaction rate at millikelvin range due to phase-locking mechanism \cite{spin_exchange2,cote_PRL}. The measured rates of the different isotopes, and in particular the variation of the even from odd isotopes of strontium might give evidence for this mechanism, and unveil the ultra-cold regime of atom-ion interactions which is otherwise inaccessible with standard ion traps \cite{cetina_heating,optical_trap}.  

Quantum-logic detection of chemical reactions can potentially be applied for various other pairs of cold chemistry neutral and ionic pairs of atoms or molecules, to calibrate their two-body potentials, determine reaction cross-sections, and observe other quantum signatures such as Feshbach or shape-type resonances \cite{optical_trap,resonances,resonances2}. Importantly, the presence of the logic ion located several micrometers away from the chemistry ion has a negligible effect on the reaction itself, safely acting as a probe that only measures the reaction outcome. 

Our scheme can also be applied to study sympathetic cooling of translational and internal state energies of molecules via collisions with ultra-cold atomic bath. Via imprint of the motion of a logic ion, the time trajectories and reaction pathways of quenching of rotational state can potentially be studied, and prepare the molecules in the ground-state. Unlike other spectroscopic methods which dissociate the molecules \cite{Brown1}, the logic method should be non-destructive.

Finally, the logic technique can be further applied to study reactions of atomic species which lack optical transitions. One example is studying the spin dependence of resonant charge-exchange reactions between a closed-shell ion (e.g.~$^{87}\text{Rb}^{+}$) and its parent atom (e.g.~$^{87}\text{Rb}$) \cite{Gao1}. In these reactions, release of hyperfine energy by Rb atoms prepared in $F=2$ can be monitored via electron-shelving detection of an additional logic ion. Alternatively, transitions of $\text{Rb}^{+}$ nonzero nuclear spin between its different magnetic levels by collisions with Rb atoms prepared in $F=1$ can potentially be monitored by side-band spectroscopy of the logic ion in the presence of a large magnetic field.

\begin{acknowledgments}
We thank Ziv Meir for usefull comments on the manuscript. This work was supported by the Israeli Science Foundation, the Israeli Ministry of Science Technology and Space and the Minerva Stiftung.
\end{acknowledgments}

\clearpage

\part*{\centerline{Methods}}

\section*{Numerical model of electron-shelving detection}
The demonstrated electron-shelving technique enables to characterize the motion of the logic ion by utilizing its coherent interaction with photons. In the presence of an optical field, the interaction Hamiltonian of the trapped ion is given by \begin{equation}
    H_I=  \tfrac{1}{2}{\Omega}\left(e^{-i\delta t}e^{i\boldsymbol{k}\cdot\boldsymbol{R}}\sigma_{+}+e^{i\delta t}e^{-i\boldsymbol{k}\cdot\boldsymbol{R}}\sigma_{-}\right).\label{eq:H_I}
\end{equation} Here $\sigma_{-}=\sigma_{+}^{\dagger}=|g\rangle\langle e|$ denotes the transition operator between a single ground state $|g\rangle$ in the S$_{1/2}$ manifold to a particular excited state $|e\rangle$ in the D$_{5/2}$ manifold. $\boldsymbol{k}$ denotes the wavenumber vector of the shelving beam, $\delta$ denotes the frequency detuning of the optical field from the atomic line, $\Omega$ denotes the Rabi angular frequency of the quadruple transition, and $\boldsymbol{R}(t)$ denotes the position of the logic ion in the trap.

For a finite shelving pulse of time $T$ starting at time $t_0$, the state of the ion evolves by the unitary operator  
\begin{equation}
    U_I =  \mathcal{T}\exp \Bigl( -i\int_{t_0}^{t_0+T}H_{I}(t')dt' \Bigr),\label{eq:U_I}
\end{equation}
where $\mathcal{T}$ denotes the time-ordering operator. Notably, the ion motion renders the Hamiltonian in Eq.~(\ref{eq:H_I}) non-commuting with itself at different times. For an ion initially in the ground state $|g\rangle$, we characterize the probability to measure a bright ion (i.e.~in the S manifold) after a single $\pi$-pulse by $P^{(1)} = \left|\langle g|U(T_{\pi})|g\rangle\right|^{2}$, where the $\pi$-pulse time $T_{\pi}=\pi/\Omega$ is determined by the laser power and the transition strength. In our experiment we used $T_{\pi}=5\,\mu$s for the first shelving pulse and $T_{\pi}=12\,\mu$s for the second.  We also consider the probability of measuring a bright ion using two independent pulses $P_\text{D}^{(2)}$ of different transitions. Here we present the case in which the effect of each pulse is similar (i.e.~ $P_\text{D}^{(2)}\approx\bigl(P^{(1)}\bigr)^2$), which is relevant in the experimentally realized limit of $T_{\pi}\omega_{z1}\gg1$, where $\omega_{z1}=2\pi\times480$ KHz is the in-phase axial motional frequency.

In between collisions, the trapped logic ion oscillates and its mean secular position follows \begin{equation}
	R_{i}(t)=\sum_{j=1}^{N_{{\mathrm{ions}}}}b_{ij}A_{ij}\cos(\omega_{ij}t+\phi_{ij}),\label{eq:r_t}
\end{equation} where $b_{ij}$ are the mode participation factors of the logic ion in the chain ($|b_{ij}|=1/\sqrt{2}$ for a two ion crystal with near equal masses), $i\in\{x,y,z\}$ index the trap axes, $j\in\{1,2\}$ index the two eigenmodes in each axis, and $\omega_{ij}$ are the trap secular frequencies of motion. In our experiment the single ion trap frequencies were $0.4$ MHz  and $1$ MHz  for the radial modes. The phases of motion $\phi_{ij}$ are randomly distributed every experimental realization whereas the amplitudes $A_{ij}$ depend on the phonons distribution in each mode.

\begin{figure*}[t]
\begin{centering}
\includegraphics[width=17.7cm]{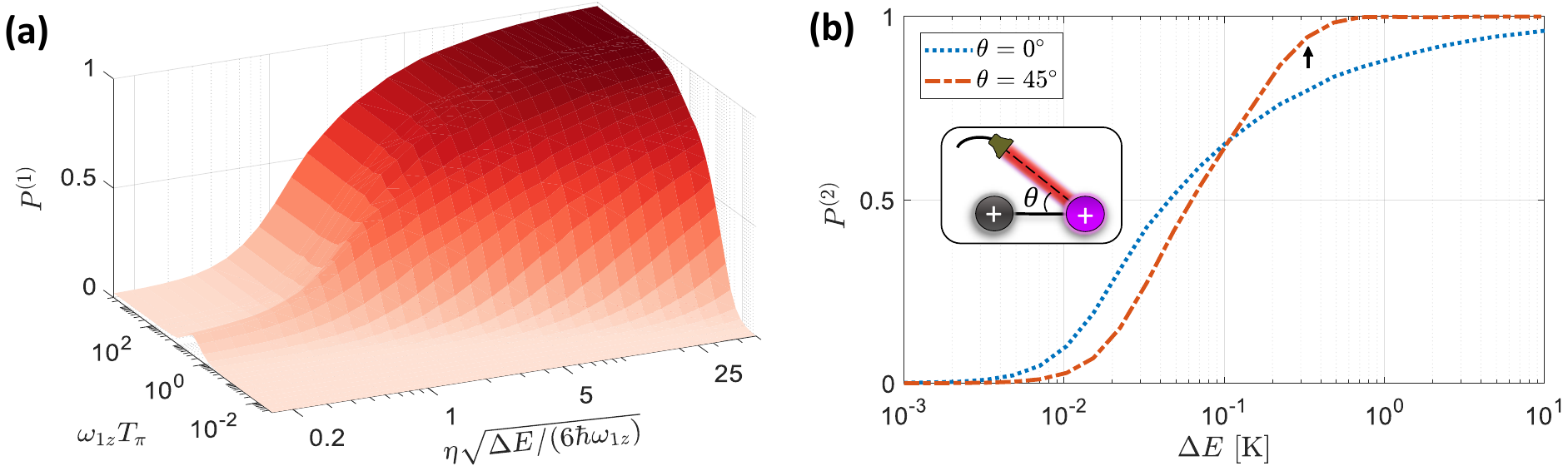}
\par\end{centering}
\centering{}\caption{\textbf{Detection efficiency of electron shelving technique.} (a) numerical calculation of the probability $P^{(1)}$ to measure the ion in the S, electronic ground-state, (bright) after a single shelving $\pi$-pulse of duration $T_{\pi}$, following a single exothermic reaction with energy $\Delta E$.  The in-phase axial trap frequency of motion of the two ion crystal is $\omega_a$ and $\eta$ is the Lamb-Dicke parameter of the shelving beam, co-linear with the trap axis ($\theta=0$). For energetic reactions ($\Delta E\gg\eta^2\hbar\omega_{1z}$) and low beam power ($\omega_{z1}T_{\pi}\gtrsim1$) the detection of hot events approaches unity. (b) Detection efficiency of two pulses of electron shelving in the low beam power limit with the experimental parameters. A tilted beam (red) which simultaneously measures radial and axial motion is more efficient than a beam colinear with the trap axis (blue). Black arrow marks the energy of hyperfine-changing collision studied in the experimental realization.\label{fig:numerical_detection_efficiency}}
\end{figure*}

To characterize the detection efficiency of this technique, we numerically simulate the dynamics of a trapped ions crystal following a single instantaneous collision with a cold neutral atom moving at a velocity $\boldsymbol{v}_{\mathrm{atom}}$ in the lab frame. Before and after the collision, the positions of the ions are evolved in the trap according to the formula in Ref.~\cite{Blatt_NJP}, taking into account both inherent and excess micromotion due to stray static fields. We assume that a collision occurs at $t_{\textrm{c}}$ and account for the effect of an exothermic reaction by updating the coefficients $A_{ij}$ and $\phi_{ij}$ in Eq.~(\ref{eq:r_t}) to new values at $t_{\textrm{c}}$. The update is determined by finding the amplitudes and phases which keep the positions $R_i(t_c)$ unchanged but vary the velocity of the colliding ion at $t_{\textrm{c}}$ in the lab frame by \begin{equation}
\boldsymbol{v}_{\mathrm{ion}}\rightarrow\left(1-r+\alpha r \mathcal{R}(\varphi)\right)(\boldsymbol{v}_{\mathrm{ion}}-\boldsymbol{v}_{\mathrm{atom}})+{v}_{\mathrm{atom}}.\label{eq:v_lab}
\end{equation} Here $\mu=m_\textrm{a} m_\textrm{i}/(m_\textrm{a}+m_\textrm{i})$ is the reduced mass, $r=\mu/m_\textrm{i}$, $\mathcal{R}$ is the rotation matrix and $\varphi$ is the scattering angle, randomly generated for Langevin-type collisions \cite{Zipkes2011}. The unit-less factor $\alpha=\sqrt{1+2r\Delta E/(m_\textrm{i}|{\bar{v}}_{\mathrm{ion}}|^2)}$ describes the increase of the ion's speed ${\bar{v}}_{\mathrm{ion}}$ in the center of mass frame, gaining kinetic energy by the exothermic reaction. 

Following the collision dynamics, we calculate the detection probabilities $P^{(1)}$ and $P^{(2)}$ by numerically solving Eq.~(\ref{eq:U_I}) with the discrete Suzuki-Trotter expansion. At this stage, we substitute the coherent evolution of the ions positions from Eq.~(\ref{eq:r_t}) in the Hamiltonian [Eq.~(\ref{eq:H_I})] for $t_0>t_c$, as we measure the ions many motional cycles after the collision time (experimentally waiting a few milliseconds after the cloud passage). As the exact trap frequencies are incommensurate, the stochastic timing of the collision  effectively randomizes the phases $\phi_{ij}$ used in the detection stage at $t_0$. We repeat the calculations and average the presented results for the random collision parameters including the scattering angle $\varphi$, initial phases $\phi_{ij}$ and initial amplitudes $A_{ij}$. The latter are randomly sampled, following a thermal distribution in a harmonic trap with an averaged temperature of 1 $k_B\times$mK. 

The electron-shelving technique is sensitive to motion along the optical axis of the shelving beam as realized by Eq.~(\ref{eq:H_I}) through the inner product between $\boldsymbol{k}$ and $\boldsymbol{R}$.  The detection efficiency depends on two type of unitless parameters $\xi_{ij}=k_{i}A_{ij}/\sqrt{2}$  and $\zeta_{ij}=\omega_{ij}T_{\pi}$. The parameters $\xi_{ij}$ determine the modulation indices of the shelving operation for each harmonic component of the ion's secular motion in the trap. It is directly associated with the number of phonons $N_{ij}$ in each mode (indexed by $ij$) and their Lamb-Dicke parameters $\eta_{ij}=k_i\sqrt{\hbar/(2m_\textrm{i}\omega_{ij})}$ by $\xi_{ij}=\eta_{ij}\sqrt{2N_{ij}}$, where in the limit that the ion motion is dominated by the reaction ($\alpha\gg1$), the average phonon number in each mode is $\langle N_{ij}\rangle=\sqrt{rr'\Delta E /(\hbar\omega_{ij})}$ where $r=m_\textrm{a}/(m_\textrm{a}+m_\textrm{i})$ and $r'\approx1/6$. The parameters $\zeta_{ij}$ quantify the number of motional cycles of each mode during the $\pi$ pulse time. 

We first discuss a simplified configuration in which the beam, tilted with an angle $\theta$ with respect to $\hat{z}$, co-aligns with the chain's axis ($\theta=0$), and then discuss the experimentally realized configuration in which $\theta=45$. In Fig.~(\ref{fig:numerical_detection_efficiency}a), we present the numerically calculated brightness of the ion $P^{(1)}$ for $\theta=0$ as a function of $\langle \xi_{z1}\rangle$ and $\zeta_{z1}$ (associated with the in-phase axial mode), for the carrier transition ($\delta=0$). Notably, this configuration is sensitive only to $\langle \xi_{zj}\rangle$ and $\zeta_{zj}$, where the averaged $j=2$ coefficients are directly determined by the $j=1$ coefficients for the two ions crystal. With sufficient number of motional oscillations within a $\pi$-pulse, the ion appears bright for energetic collisions, whereas it is efficiently shelved to the D manifold and appears dark otherwise. In the low laser power limit $(\zeta_{zj}\gg1)$, the modulated spectrum of the ion is spectrally resolved, and the phase modulation is manifested as a reduction of the Rabi frequency by the product of zeroth order Bessel functions $\Pi_{ij} J_0(\xi_{ij})$ \cite{Wineland_1998}, yielding $P^{(1)}= \cos^2\left(\Pi_{ij} J_0(\xi_{ij})\right)$ for narrow linewidth transitions.
In Fig.~(\ref{fig:numerical_detection_efficiency}b), we present the probability of double-shelving for $\theta=45^{\circ}$ as realized experimentally, compared to the co-aligned case ($\theta=0^{\circ}$). Evidently, the detection probability as a function of collision energy sharpens for nonzero $\theta$, due to contributions of additional nonzero $\xi_{ij}$ that reduce the Rabi frequency of the hot ion. In black arrow we indicate the predicted probability to detect a single hyperfine-changing reaction in the experiment.

It is insightful to compare the present detection method with monitoring of Doppler cooling curves \cite{Ozeri_2007} and discuss its possible utilization in precision thermonetry. During Doppler cooling, a hot ion scatters photons less efficiently due to the motion-induced phase modulation. Monitoring the number of scattered photons as a function of time enables to infer the ion's temperature. Both Doppler cooling and electron-shelving methods detect the motion via the phase modulation mechanism which imparts the optical excitation of a hot ion with respect to a cold one. In the optical Bloch sphere picture, the mechanism corresponds to mapping the phase of the complex Rabi frequency to the rotation axis of the atom by the beam; a cold ion has a stationary phase and a static axis whereas a hot ion with a fluctuating phase would lead to time-varying rotation axis and inefficient transfer from the south to north pole. However, Doppler cooling is often practically imparted by the photon collection efficiency of the detection. For low collection efficiencies, observation of the motion-dependent fluorescence requires scattering of numerous photons, which in turn cool the ion and render the detection less sensitive. Consequently, the motion-dependent signal has to overcome the photon shot noise at the detection stage, often limiting the sensitivity of a single event to temperatures above 1 Kelvin \cite{Ziv_sys}. In contrast, the electron shelving requires absorption of a single photon to map the motional state of the ion to its electronic state, which can later be detected via fluorescence in a non-destructive manner. This method can potentially be utilized as a thermometery and enable detection of $\Delta E$, e.g.~by varying $T_{\pi}$ or trap frequencies, and calculating the likelihood with respect to the model.

\clearpage
\setcounter{equation}{0}
\setcounter{figure}{0}
\setcounter{table}{0}

\def\equationautorefname~#1\null{Equation (#1)\null}
\newcommand{\diff}{\mathrm{d}}

\renewcommand{\theequation}{S\arabic{equation}}
\renewcommand{\thefigure}{S\arabic{figure}}
\renewcommand{\thetable}{S-\Roman{table}}

\onecolumngrid 
\setcounter{page}{1}

\part*{\centerline{Supplementary Information}}

\section*{Neutral atoms setup}
The upper vacuum chamber of our setup serves for neutral atoms preparation. We laser cool $^{87}$Rb vapor in a Magneto Optical Trap (MOT) using six optical cooling beams, an optical repump beam and a set of anti-Helmholtz coils. The cooling beams originate from a Distributed Feedback (DFB) laser at $780$ nm, which is frequency locked to a reference laser which is locked to the atomic line in a Doppler-free saturation absorption spectroscopy. The cooling beams are power amplified via a tapered amplifier and pass an acousto-optic modulator (AOM) which red-detune them by $20$ MHz from the $F=2\rightarrow F'=3$ ($D2$) cycling transition. The beams polarizations are set circular and each port has about 20 mW with a typical waist of about $2.2$ cm. 
The repump beam originates from a different $780$ nm DFB laser whose frequency is locked near the $F=1\rightarrow2$ of the $D2$ transition via additional absorption spectroscopy cell,  emptying the lower hyperfine state which is transparent to the cooling light. The power of this beam is few mW, and is retro reflected through the cell to double its rate. A vapor of rubidium is maintained via continuous heating of an isotopically enriched $^{87}$Rb dispenser and illumination of the upper chamber walls with multiple UV-lamps during the loading stage. These lamps stimulate light-induced desorption of rubidium sticked to the inner surface of the chamber and efficiently increase the rubidium pressure. To apply optical restraining forces for the trapping, we apply a magnetic gradient of about $10$G/cm during the initial loading stage. 

We realized the following cooling sequence which we found experimentally efficient. We Doppler cooled and compressed the atomic cloud for about $90$ ms, by gradually lowering the beams power. We then realized polarization gradient cooling (PGC) by turning off the magnetic gradient field and zeroing the background magnetic field. This stage was followed by gradual turn-off of the optical beams in a $5$ ms window with simultaneous red-detunning of the cooling beams further away from the atomic optical resonance. This stage resulted with an atomic cloud at a temperature of a few $\mu$K as measured by time of flight (TOF). 

An atomic cloud of about $10^6$ atoms was loaded into an off-resonant optical lattice dipole trap, which is formed by continuously-running two counter propagating beams. The beams originated from an amplified YAG laser at $1064$ nm passed through two different double-pass AOMs which enabled to vary their optical frequencies. The beams (about $1.5$W each with an estimated waist of about $220\,\mu\textrm{m}$) were aligned to pass through the center of the Paul trap (by maximizing their induced light-shift on the ions) and near the center of the atomic cloud. The power of the beams was constantly regulated to avoid light-shift drifts of the ion's optical frequency. Control of the magnetic field at the cooling stage preceding PGC enabled to position and spatially overlap the center of the MOT with the optical-dipole trap. 

We initialize the spin state of the atoms with a sequence of optical pumping and microwave pulses. First, we optically-pumped the atoms to the lower, $F=1$ hyperfine state using a linearly polarized $780$ nm beam which originated from a different DFB laser tuned near the $F=2\rightarrow F'=1$.  Next, we pumped the atoms to either one of the $|1,\pm1\rangle$ states using a sequence of $\pi$-pulses of microwave field which originated from an impedance-matched antenna near the hypefine frequency of $^{87}$Rb at $6.8$ GHz. A static magnetic field of $0.61$ G enabled to spectrally resolve the magnetic transitions. We found it efficient to  use several microwave pulses resonant near the $|1,0\rangle\rightarrow|2,\pm1\rangle$ and  $|1,\mp1\rangle\rightarrow|2,0\rangle$ magnetic transitions, followed by optical pumping pulses which deplete the upper hyperfine manifold. Finally we transferred the atoms to any of the $|2,m\rangle $ by adiabatically ramping the microwave field across the $|1,\pm1\rangle\rightarrow|2, |m|\rangle $ transition (for $m=-2,-1,0,1,2$). This coherent transfer was found efficient for transferring the entire cloud population despite the presence of a residual magnetic gradient, which induced space-dependent variation of the magnetic resonance frequency within the cloud.

We shuttled the atomic cloud to the lower chamber by chirping the relative
optical frequencies of the two optical lattice beams. The cold atomic cloud adiabatically followed the moving optical trap in a trajectory similar to that presented in Ref.~\cite{BS2021}, reaching the ion trap in about 100 msec at a mean velocity of $\langle v\rangle=248.6$ cm/s. The atoms are decelerated before passing the ions to a velocity of $v_{\rm a}=23.8$ cm/s. We verified that the spin state is preserved during shuttling, and that shuttling-induced atom loss is small.

\section*{Ions setup}
We load and trap Sr$^{+}$ ions in a linear Paul trap. The trap is constructed from segmented titanium blades driven by a time-dependent radio-frequency (RF) and static potentials. The RF potential oscillates at $26.5$ MHz and is amplified in a home-made helical resonator. The trap generates time-dependent electric field lines in the radial, $xy$ plane, which exert ponder-motive radial trapping forces with secular single-ion trap frequencies of about $f_x=400$ kHz and $f_y=1$ MHz. The static potential produces static axial field along $z$, corresponding to an harmonic trap with frequency $f_z=480$ KHz. 
 
We load individual ions into the trap via photoionization of neutral strontium vapor. A home-made strontium oven is heated beneath the trap, emitting a hot jet of atoms. Two counter-propagating beams at $461$ nm and $405$ nm ionize the neutral strontium atom via a two photon transition, first exciting the transition $S\rightarrow P$, and then to the continuum. The $461$ light originates from an External Cavity Diode Laser (ECDL) at $923$ nm which is tapered amplified and frequency doubled. The ECDL is frequency locked to a wavemeter while the $405$ nm diode is free running. Isotope selectivity in the loading is achieved by setting the $461$ nm light in resonance with the neutral atom line including the isotope frequency shifts \cite{shift_neutral} and maintaining the atomic absorption line narrow. The latter is realized by keeping the beams power low and aligning their wavevectors transverse to the atomic motion to suppress Doppler broadening. 

To trap a hot photo-ionized atom efficiently, we Doppler cool it with $422$ nm light red-tuned by about 230 MHz from the ion's $\text{S}_{1/2}\rightarrow \text{P}_{1/2}$ resonance. Simultaneously, we use $1092$ nm co-propagating light resonant with the $\text{D}_{3/2}\rightarrow \text{P}_{3/2}$ line to deplete the long-lived D state and maintain the cooling. Importantly we account for the ion's isotope shift during this initial cooling stage \cite{shift_ion}, which is particularly important for trapping of the low abundant ($0.56\%$) $^{84}\text{Sr}^{+}$ isotope. The $422$ nm originates from a diode laser which is injected-locked to reference lasers. The $422$ nm reference light originates from an ECDL at $843$ nm which is frequency doubled and uses double frequency-locking stages. One lock controls the laser's current using a Pound-Driver-Hall (PDH) error signal from an optical cavity whereas a second lock controls the grating's piezo using an error signal from a saturation absorption Doppler-free spectroscopy of the $5\text{S}_{1/2} (F=2)\rightarrow 6\text{P}_{1/2}(F=3)$ transition of rubidium vapor. The reference $1092$ nm light is locked to an optical cavity using PDH error signal.

To enable coherent manipulation and state-dependent detection of the logic ion, we used $674$ nm light tuned near the $\text{S}_{1/2}\rightarrow\text{D}_{5/2}$ narrow quadruple transition, and $1033$ nm light resonant with the $\text{D}_{5/2}\rightarrow \text{P}_{3/2}$ transition. The beams pass through AOMs to allow for frequency control and are co-aligned with the Doppler-cooling beam near the lower chamber, propagating at $45^{\circ}$ with respect to $\hat{z}$ at the center of the trap, and has overlap with all trap modes. The $674$ nm light originates from a diode which is injection locked using a narrow reference laser. The reference laser is frequency locked and narrowed by a ultra-high-finesse optical cavity using a PDH error signal, resulting with a $\Delta f\approx20$ Hz linewidth. Tuning the optical frequency near different transitions in the form $|m,n\rangle\rightarrow |m',n'\rangle$ enables to change the z-projection of the spin from $m$ in the S manifold to $m'$ in the D manifold, as well as to change the number of phonons in one of the motional modes from $n$ to $n'$. A constant $3$ G magnetic field enabled to optically resolve the different transitions.

State preparation of the ions is performed during the shuttling time of the atoms towards the lower chamber. We Doppler cooled the logic ion using a 422 beam tuned near the optical resonance, using a narrow dark resonance. The $422$ nm and $1092$ nm powers and detunings were optimized to improve the cooling of the logic ion. To ensure that the chemistry ion ends in the ground state (i.e.~not accidentally excited to the long-lived $\text{D}_{3/2}$ state due to off-resonant absorption), we phase-modulated the $1092$ nm light and tuned its side-bands on resonance with the isotope-shifted transitions of the chemistry ion. A second $422$ nm off resonant beam (used for trapping) was simultaneously applied, efficiently cooling rare events in which the ions were hot. We then performed side band cooling (SBC) for the six motional modes of the crystal, combined with optical pumping (OP) of the logic ion. For an up-polarized logic ion, we used the transition $|\tfrac{1}{2},n\rangle\rightarrow |\tfrac{5}{2},n-1\rangle$ for SBC of each of the modes and the transition $|-\tfrac{1}{2},n\rangle\rightarrow |\tfrac{3}{2},n\rangle$ for OP, followed by $1033$ nm pulses to quickly deplete the $\text{D}_{5/2}$ state. 
State detection of the logic ion is performed with two shelving pulses as described in the Methods section.

The combined atom-ion sequence was repeated several thousands of times at each measurement point, with a repetition rate of about 3 Hz. Every 250 shots we imaged the ions to ensure normal operation, and every few thousands measurements imaged the atomic cloud, using absorption imaging, to account and correct for small drifts of its density. Every few hours we performed calibration and compensation protocols. We calibrated for the
the Rabi frequency of the 674 light, which was typically around $100$ kHz and for small variations of the 674-optical transitions which could result from drift in the magnetic field or light shift.

We also tracked the frequencies of the six motional modes, verifying the mass induced shift. For logic and chemistry ions whose mass difference $\delta m=m_{\rm C}-m_{\rm L}$ is small ($|\delta m|\ll m_{\rm L}$) the axial center-of-mass frequency is shifted by a factor of $[1-\delta m/(4m_{\rm L})]$ with respect to the frequency of a single logic ion as shown in Fig.~\ref{fig:setup} (black dashed lines). 

We also minimized Excess Micro-Motion (EMM) which originate from stray fields using two sets of DC electrodes and additional RF electrodes. The compensation voltages were determined by minimization of the transition probability of a long-duration $674$ nm light pulse tuned near the RF ($26.5$ MHz) side-band using three spatially different beams. We estimate that collision happened at a temperature of about 1 mK.



\begin{thebibliography}{10}

\bibitem{qchem} A. Szabo and N. Ostlund 1996 "Modern Quantum Chemistry" (New York: Dover).

\bibitem{CompPow} R. A. Friesner, "Ab initio quantum chemistry: Methodology and applications". PNAS, 102 (19) 6648-6653; (2005).

\bibitem{CompPow2} D. S. Abrams, and S. Lloyd, "Quantum Algorithm Providing Exponential Speed Increase for Finding Eigenvalues and Eigenvectors". Phys. Rev. Lett. 83, 5162 (1999).


\bibitem{atoms_Feshbach} C. Chin, R. Grimm, P. S. Julienne, and E. Tiesinga, "Feshbach resonances in ultracold gases", Rev. Mod. Phys. 82, 1225 (2010).

\bibitem{atoms_Feshbach2} T. K\"ohler, K. G\'oral, and P. S. Julienne, "Production of cold molecules via magnetically tunable Feshbach resonances", Rev. Mod. Phys. 78, 1311 (2006).

\bibitem{atoms_shape} P. Paliwal, N. Deb, D. M. Reich, A. van der Avoird, C. P. Koch, and E. Narevicius, "Determining the nature of quantum resonances by probing elastic and reactive scattering in cold collisions" Nat. Chem. 13, 94-98 (2021).


\bibitem{hybrid_RMP} M. Tomza, K. Jachymski, R. Gerritsma, A. Negretti, T. Calarco, Z. Idziaszek, and P. S. Julienne, "Cold hybrid ion-atom systems". Rev. Mod. Phys. 91, 035001 (2019).

\bibitem{Nature_Kohl} C. Zipkes, S. Palzer, C. Sias, and M. K\"ohl, "A trapped single ion inside a Bose-Einstein condensate". Nature 464, 388-391 (2010). 


\bibitem{hybrid_RMP2} A. H\"arter and J. Hecker Denschlag, "Cold atom-ion experiments in hybrid traps", Contemp. Phys. 55, 33 (2014).

\bibitem{hybrid_RMP3} W. Stefan. "Ion-atom hybrid systems". Proceedings of the International School of Physics "Enrico Fermi", 189. pp. 255-268.(2015)

\bibitem{Ziv_sys} Z. Meir, T. Sikorsky, R. Ben-shlomi, N. Akerman, M. Pinkas, Y. Dallal, and R. Ozeri, "Experimental apparatus for overlapping a ground-state cooled ion with ultracold atoms". J. Mod. Opt. 65, 501 (2018).

\bibitem{spin_exchange1}T. Sikorsky, Z. Meir, R. Ben-shlomi, N. Akerman and R. Ozeri, "Spin-controlled atom-ion chemistry". Nat. Commun. 9, 920 (2018).

\bibitem{spin_exchange2} T. Sikorsky, M. Morita, Z. Meir, A. A. Buchachenko, R. Ben-shlomi, N. Akerman, E. Narevicius, T. V. Tscherbul, and R. Ozeri, "Phase Locking between Different Partial Waves in Atom-Ion Spin-Exchange Collisions". Phys. Rev. Lett. 121, 173402 (2018).

\bibitem{spin_exchange3}
F. G. Major and H. G. Dehmelt. "Exchange-Collision Technique for the rf Spectroscopy of Stored Ions". Phys. Rev. 170, 91 (1968).

\bibitem{optical_trap} T. Feldker, H. F\"urst, H. Hirzler, N. V. Ewald, M. Mazzanti, D. Wiater, M. Tomza, and R. Gerritsma, "Buffer gas cooling of a trapped ion to the quantum regime". Nat. Phys. 16, 413-416 (2020).

\bibitem{spin_relaxation1}L. Ratschbacher, C. Sias, L. Carcagni, J. M. Silver, C. Zipkes, and M. K\"ohl, "Decoherence of a Single-Ion Qubit Immersed in a Spin-Polarized Atomic Bath". Phys. Rev. Lett. 110, 160402 (2013).

\bibitem{spin_relaxation2} T. V. Tscherbul, P. Brumer, and A. A. Buchachenko, "Spin-Orbit Interactions and Quantum Spin Dynamics in Cold Ion-Atom Collisions". Phys. Rev. Lett. 117, 143201 (2016).

\bibitem{charge_exchange0} L. Ratschbacher, C. Zipkes, C. Sias and M. K\"ohl, "Controlling chemical reactions of a single particle". Nat. Phys., 8, 649-652 (2012).


\bibitem{charge_exchange1} A. T. Grier, M. Cetina, F. Oru{\v c}evi\'c, and Vladan Vuleti\'c, "Observation of Cold Collisions between Trapped Ions and Trapped Atoms", Phys. Rev. Lett. 102, 223201 (2009).


\bibitem{charge_exchange2} W. G. Rellergert, S. T. Sullivan, S. Kotochigova, A. Petrov, K. Chen, S. J. Schowalter, and E. R. Hudson, "Measurement of a Large Chemical Reaction Rate between Ultracold Closed-Shell $^{40}\text{Ca}$ Atoms and Open-Shell $^{174}\text{Yb}^{+}$ Ions Held in a Hybrid Atom-Ion Trap". Phys. Rev. Lett. 107, 243201 (2011).

\bibitem{charge_exchange3} K. Ravi, S. Lee, A. Sharma, G. Werth, and S.A. Rangwala, "Cooling and stabilization by collisions in a mixed ion-atom system". Nat. Commun. 3, 1126 (2012).

\bibitem{charge_exchange4} H. Li, S. Jyothi, M. Li, J. Klos, A. Petrov, K. R. Brown, and S. Kotochigova, "Photon-mediated charge-exchange reactions between 39K atoms and 40Ca+ ions in a hybrid trap". Phys. Chem. Chem. Phys., 22, 10870-10881 (2020).

\bibitem{charge_exchange5} A. Mahdian, A. Krükow, J. H. Denschlag, "Direct observation of swap cooling in atom-ion collisions". arXiv preprint, arXiv:2012.07759 (2020).

\bibitem{charge_exchange6} N. V. Ewald, T. Feldker, H. Hirzler, H. A. F\"urst, and R. Gerritsma, "Observation of Interactions between Trapped Ions and Ultracold Rydberg Atoms", Phys. Rev. Lett. 122, 253401 (2019). 


\bibitem{Molecular_formation1}F. H. J. Hall, M. Aymar, N. Bouloufa-Maafa, O. Dulieu, and S. Willitsch, "Light-Assisted Ion-Neutral Reactive Processes in the Cold Regime: Radiative Molecule Formation versus Charge Exchange". Phys. Rev. Lett. 107, 243202 (2011).

\bibitem{Molecular_formation2}A. Mohammadi, A. Kr\"ukow, A. Mahdian, M. Dei{\ss}, J> P\'erez-R\'ios, H. da Silva Jr., M. Raoult, O. Dulieu, and J. Hecker Denschlag, "Life and death of a cold BaRb$^{+}$ molecule inside an ultracold cloud of Rb atoms", Phys. Rev. Research 3, 013196 (2021).

\bibitem{elastic1} S. Haze, M. Sasakawa, R. Saito, R. Nakai, and T. Mukaiyama. "Cooling Dynamics of a Single Trapped Ion via Elastic Collisions with Small-Mass Atoms". Phys. Rev. Lett. 120, 043401 (2018).

\bibitem{elastic2}C. Zipkes, S. Palzer, L. Ratschbacher, C. Sias, and M. K\"ohl, Cold Heteronuclear "Atom-Ion Collisions". Phys. Rev. Lett. 105, 133201 (2010).


\bibitem{nonequilib_1} Z. Meir, T. Sikorsky, R. Ben-shlomi, N. Akerman, Y. Dallal, and R. Ozeri, "Dynamics of a Ground-State Cooled Ion Colliding with Ultracold Atoms". Phys. Rev. Lett. 117, 243401 (2016).

\bibitem{nonequilib_2} Z. Meir, M. Pinkas, T. Sikorsky, R. Ben-shlomi, N. Akerman, and R. Ozeri, "Direct Observation of Atom-Ion Nonequilibrium Sympathetic Cooling". Phys. Rev. Lett. 121, 053402 (2018).

\bibitem{nonequilib_3} R. G. DeVoe, "Power-Law Distributions for a Trapped Ion Interacting with a Classical Buffer Gas". Phys. Rev. Lett. 102, 063001 (2009). 

\bibitem{nonequilib_4} K. Chen, S. T. Sullivan, and E. R. Hudson, "Neutral Gas Sympathetic Cooling of an Ion in a Paul Trap". Phys. Rev. Lett. 112, 143009 (2014). 

\bibitem{q_logic1} P. O. Schmidt, T. Rosenband, C. Langer, W. M. Itano, J. C. Bergquist and D. J. Wineland, "Spectroscopy Using Quantum Logic". Science 309, 5735, 749-752 (2005).

\bibitem{q_logic2} F. Wolf, Y. Wan, J. C. Heip, F. Gebert, C. Shi and P. O. Schmidt, "Non-destructive state detection for quantum logic spectroscopy of molecular ions". Nature 530, 457-460 (2016).

\bibitem{q_logic3} M. Sinhal, Z. Meir, K. Najafian, G. Hegi, S. Willitsch, "Quantum-nondemolition state detection and spectroscopy of single trapped molecules". Science 367, 6483, 1213-1218 (2020).

\bibitem{q_logic4} Y. Lin, D. R. Leibrandt, D. Leibfried, and C. Chou, "Quantum entanglement between an atom and a molecule". Nature volume 581, pages273-277 (2020).


\bibitem{q_logic5} C. W. Chou, A. L. Collopy, C. Kurz, Y. Lin, M. E. Harding, P. N. Plessow, T. Fortier, S. Diddams, D. Leibfried and D. R. Leibrandt, "Frequency-comb spectroscopy on pure quantum states of a single molecular ion". Science, 367 6485, 1458-1461 (2020).


\bibitem{q_logic_precision} S. M. Brewer, J.-S. Chen, A. M. Hankin, E. R. Clements, C. W. Chou, D. J. Wineland, D. B. Hume, and D. R. Leibrandt, $^{27}\text{Al}^{+}$ "Quantum-Logic Clock with a Systematic Uncertainty below $10^{-18}$". Phys. Rev. Lett. 123, 033201 (2019).


\bibitem{q_logic_precision2} F. Gebert, Y. Wan, F. Wolf, C. N. Angstmann, J. C. Berengut, P. O. Schmid, "Precision isotope shift measurements in calcium ions using quantum logic detection schemes". Phys. Rev. Lett. 115, 053003 (2015).

\bibitem{Leibfried1} D. Kienzler, Y. Wan, S. D. Erickson, J. J. Wu, A. C. Wilson, D. J. Wineland, and D. Leibfried, "Quantum Logic Spectroscopy with Ions in Thermal Motion",Phys. Rev. X 10, 021012  (2020).


\bibitem{q_logic_SM} P. Micke, T. Leopold, S. A. King, E. Benkler, L. J. Spie{\ss}, L. Schm\"oger, M. Schwarz, J. R. Crespo L\'opez-Urrutia, and P. O. Schmidt, "Coherent laser spectroscopy of highly charged ions using quantum logic". Nature 578, 60-65 (2020).

\bibitem{willitsch1} F. H. J. Hall and S. Willitsch, "Millikelvin Reactive Collisions between Sympathetically Cooled Molecular Ions and Laser-Cooled Atoms in an Ion-Atom Hybrid Trap". Phys. Rev. Lett. 109, 233202 (2012).

\bibitem{willitsch1b} K. Najafian, Z. Meir, M. Sinhal and S. Willitsch, "Identification of molecular quantum states using phase-sensitive forces", Nat. Commun. 11, 4470 (2020). 

\bibitem{willitsch2} Z. Meir, G. Hegi, K. Najafian, M. Sinhala  and  S. Willitsch, "State-selective coherent motional excitation as a new approach for the manipulation, spectroscopy and state-to-state chemistry of single molecular ions". Faraday Discuss. 217, 561-583 (2019).

\bibitem{q_logic_chem} M. Tomza and M. Lisaj, "Interactions and charge-transfer dynamics of an Al$^{+}$ ion immersed in ultracold Rb and Sr atoms". Phys. Rev. A 101, 012705 (2020).

\bibitem{cetina_heating} M. Cetina, A. T. Grier, and V. Vuleti\'c, "Micromotion-Induced Limit to Atom-Ion Sympathetic Cooling in Paul Traps". Phys. Rev. Lett. 109, 253201 (2012).

\bibitem{atom_ion_cote} R. C{\^ o}t\'e, "Ultracold Hybrid Atom-Ion Systems". Advances In Atomic, Molecular, and Optical Physics 65, 67-126 (2016).


\bibitem{Ozeri_2007} J. H. Wesenberg, R. J. Epstein, D. Leibfried, R. B. Blakestad, J. Britton, J. P. Home, W. M. Itano, J. D. Jost, E. Knill, C. Langer, R. Ozeri, S. Seidelin, and D. J. Wineland. "Fluorescence during Doppler cooling of a single trapped atom". Phys. Rev. A 76, 053416 (2007).

\bibitem{Drewsen1} M. Drewsen, A. Mortensen, R. Martinussen, P. Staanum, and J. L. S\o{}rensen, "Nondestructive Identification of Cold and Extremely Localized Single Molecular Ions". Phys. Rev. Lett. 93, 243201 (2004).


\bibitem{Pinkas2020} M. Pinkas, Z. Meir, T. Sikorsky, R. Ben-Shlomi, N. Akerman, R. Ozeri, "Effect of ion-trap parameters on energy distributions of ultra-cold
atom-ion mixtures", New Journal of Physics. 22, 013047, (2020).

\bibitem{cote_PRL} R. C{\^ o}t\'e and I. Simbotin, "Signature of the $s$-Wave Regime High above Ultralow Temperatures". Phys. Rev. Lett. 121, 173401 (2018).


\bibitem{resonances} M. Tomza, C. P. Koch, and R. Moszynski, "Cold interactions between an Yb$^{+}$ ion and a Li atom: Prospects for sympathetic cooling, radiative association, and Feshbach resonances". Phys. Rev. A 91, 042706 (2015).

\bibitem{resonances2} P. Weckesser, F. Thielemann, D. Wiater, A. Wojciechowska, L. Karpa, K. Jachymski, M. Tomza, T. Walker, T. Schaetz. "Observation of Feshbach resonances between a single ion and ultracold atoms". arXiv preprint, arXiv:2105.09382 (2021). 

\bibitem{Brown1} R. Rugango, A. T. Calvin, S. Janardan, G. Shu, and K. R. Brown, "Vibronic spectroscopy of sympathetically cooled CaH$^+$", Chem. Phys. Chem. 17, 3764 (2016).

\bibitem{Gao1} M. Li, L. You, and B. Gao, "Multichannel quantum-defect theory for ion-atom interactions". Phys. Rev. A 89, 052704 (2014).

\bibitem{Blatt_NJP} A. Bermudez, P. Schindler, T. Monz, R. Blatt, and M. M\"uller, "Micromotion-enabled improvement of quantum logic gates with trapped ions". New J. Phys. 19 113038 (2017).

\bibitem{Zipkes2011} C. Zipkes, L. Ratschbacher, S. Palzer, C. Sias and M. K\"ohl, "Hybrid quantum systems of atoms and ions".  J. Phys.: Conf. Ser. 264 012019 (2011).

\bibitem{Wineland_1998} D. J. Berkeland, J. D. Miller, J. C. Bergquist, W. M. Itano, and D. J. Wineland, "Minimization of ion micromotion in a Paul trap". Journal of Applied Physics 83, 5025 (1998).

\bibitem{BS2021} R. Ben-shlomi, M. Pinkas, Z. Meir, T. Sikorsky, O. Katz, N. Akerman, and R. Ozeri. "High-energy-resolution measurements of an ultracold-atom-ion collisional cross section". Phys. Rev. A 103, 032805 (2021).

\bibitem{shift_neutral} H. Miyake, N. C. Pisenti, P. K. Elgee, A. Sitaram, and G. K. Campbell, "Isotope-shift spectroscopy of the $^1S_0\rightarrow^3P_1$ and $^1S_0\rightarrow^3P_0$ transitions in strontium". Phys. Rev. Research 1, 033113 (2019).

\bibitem{shift_ion} B. Dubost, R. Dubessy, B. Szymanski, S. Guibal, J.-P. Likforman, and L. Guidoni, "Isotope shifts of natural Sr$^+$ measured by laser fluorescence in a sympathetically cooled Coulomb crystal". Phys. Rev. A 89, 032504 (2014).

\end{thebibliography}
\end{document}